\documentclass[fleqn,twoside]{article}
\usepackage{espcrc2,amsmath,epsfig}

\newcommand{\rf}[4]{{#1} {\bf #2}, #3 (#4)}

\newcommand{\pr}{Phys.\ Rev.}
\newcommand{\physrl}{Phys.\ Rev.\ Lett.}
\newcommand{\physl}{Phys.\ Lett.}

\newcommand{\npps}[3]{Nucl.\ Phys. {\bf B} (Proc.\ Suppl.) #1, 
	#2 (#3)}

\newcommand{\lapI}{$\partial^2$(I) }
\newcommand{\lapII}{$\partial^2$(II) }
\newcommand{\lapIII}{$\partial^2$(III) }

\newcommand{\oa}[1]{\ensuremath{{\cal O}(a^{#1})}}
\newcommand{\oag}[2]{\ensuremath{{\cal O}(a^{#1}g^{#2})}}

\newcommand{\eref}[1]{Eq.~(\ref{#1})}

\title{Quark propagator from an improved staggered action in Laplacian and
       Landau gauges}
 
\author{Patrick O.~Bowman\address[FSU]{Department of Physics and CSIT, 
Florida State University, Tallahassee FL 32306-4120, USA}\thanks{Presented by
POB.},
Urs M.~Heller\addressmark[FSU] and
Anthony G.~Williams\address[ADL]{CSSM and 
Department of Physics and Mathematical Physics,
Adelaide University, Australia 5005}}

\begin{document}

\begin{abstract}
Studies of gauge dependent quantities are afflicted with Gribov copies, but
Laplacian gauge fixing provides one possible solution to this problem.
We present results for the lattice quark propagator in both Landau and 
Laplacian gauges using standard and improved staggered quark actions.
The standard Kogut-Susskind action has errors of \oa{2} while the improved 
``Asqtad'' action has \oa{4}, \oag{2}{2} errors and this improvement is seen 
in the quark propagator.  We demonstrate the application of 
tree-level corrections to these actions and see that Landau and Laplacian 
gauges produce very similar results.  In addition, we test an ansatz for the
quark mass function, with promising results.  In the chiral limit, the 
infrared quark mass, $M(q^2 = 0)$ is found to be $260\pm 20$ MeV. 
\end{abstract}

\maketitle

\section{The quark propagator}

The quark propagator is a fundamental quantity of QCD.  Though gauge 
dependent, it manifestly displays dynamical chiral symmetry breaking, 
contains the chiral condensate and $\Lambda_{\text{QCD}}$, and has
been used to compute the running quark mass~\cite{Aok99,Bec00b}.  Some model 
hadron calculations rely on ans\"{a}tze for the quark propagator, yet 
on the lattice we have the opportunity to study it in a direct,
nonperturbative fashion.  Quark propagator studies can be complicated,
however, by strong lattice artefacts~\cite{Sku01a,Sku01b}; it has also been
studied using the overlap action~\cite{Bon01b}.

We are required to fix a gauge and we choose the Landau and 
the Laplacian gauges~\cite{Lapgag}.  We use Wilson glue at 
$\beta = 5.85$ ($a \simeq 0.125$ fm) on a 
$16^3 \times 32$ lattice and six quark masses from 
$am = 0.075$ down to 0.0125 (115 to 19 MeV).  
Calculations were done on 80 configurations.

We use the standard Kogut-Susskind (KS) action and the ``Asqtad'' quark 
action~\cite{Org99}, a fat-link staggered action 
that combines three-link, five-link and seven-link staples to minimise flavour
changing interactions along with the three-link Naik
term and planar five-link Lepage term.  The 
coefficients are tadpole improved and chosen to remove all tree-level \oa{2} 
errors.  This action was motivated by the desire to improve flavour symmetry, 
but has also been reported to have good rotational properties and small
mass renormalisation~\cite{Hei01}.

In the (Euclidean) continuum, Lorentz invariance allows us to decompose the 
full quark propagator into Dirac vector and scalar pieces
\begin{equation}
S^{-1}(p^2) = Z^{-1}(p^2) [i \gamma \cdot p + M(p^2)].
\end{equation}
Asymptotic freedom means that, as $p^2 \rightarrow \infty$, 
$S^{-1}(p^2) \rightarrow  i\gamma \cdot p + m,$ (the free propagator)
where $m$ is the bare quark mass.

From consideration of the tree-level forms of our two lattice actions, we
define the momentum variables $q_\mu \equiv \sin(p_\mu)$ for the KS
action and 
\begin{equation}
q_\mu \equiv \sin(p_\mu) \bigl[ 1 + \frac{1}{6} \sin^2(p_\mu) \bigr]
\end{equation}
for the Asqtad action, where $p_\mu$ is the usual lattice momentum,
\begin{equation}
p_\mu = \frac{2\pi n_\mu}{aL_\mu} \qquad n_\mu \in 
	\Bigl[ \frac{-L_\mu}{4}, \frac{L_\mu}{4} \Bigr).
\end{equation}
By considering the propagator as a function of $q_\mu$ instead of $p_\mu$, we 
ensure that the lattice quark propagator has the correct tree-level form 
and hopefully better approximates its continuum behaviour.  This is the same
philosphy that has been used in studies of the gluon propagator (see 
Ref.~\cite{Bon01} and references therein).  
See also footnote 6 in Ref.~\cite{Bec00b}.

To help us identify lattice artefacts - as we are employing only one set of
configurations - we separate the data according to the direction in which the
momentum lies.  Data from momenta lying wholly on a spatial cartesian 
direction are plotted with squares, along the temporal direction, triangles, 
and along the four-diagonal, with diamonds.  Circles represent any other 
combination.

\section{Laplacian gauge}

Laplacian gauge is a nonlinear gauge fixing that respects rotational
invariance yet is free of Gribov ambiguities.  Although it is 
difficult to understand perturbatively, it is equivalent to Landau gauge 
in the asymptotic region~\cite{vBa95,Man01}.  
It is also computationally cheaper then Landau gauge.  There is, however, more
than one way of obtaining such a gauge fixing in $SU(N)$. 
The three implementations of Laplacian gauge fixing used here are:
\begin{enumerate}
\item \lapI gauge (QR decomposition), used in Ref.~\cite{Ale01}.
\item \lapII gauge, where the complex 3x3 matrix is projected 
	onto $SU(3)$ by maximising its trace.  This will be discussed in more
	detail in an upcoming paper on the gluon propagator~\cite{Bow01b}. 
\item \lapIII gauge (Polar decomposition), the original prescription described
 	in Ref.~\cite{Lapgag}.
\end{enumerate}

The gauge transformations employed in Laplacian gauge fixing are constructed 
from the lowest eigenvectors of the covariant lattice Laplacian operator.  The
way that these eigenvectors transform under gauge transformations provides the
uniqueness of the Laplacian gauge.
The three implementations discussed differ in the way that the gauge 
transformation is constructed from those eigenvectors.  We can 
think of each projection method as defining its own Laplacian 
gauge.  In all cases the resulting gauge is unambiguous for all 
configurations except a set of measure zero.

\section{Results}

\subsection{Performance of the Asqtad action}

\begin{figure}[bt]
\begin{center}
\epsfig{figure=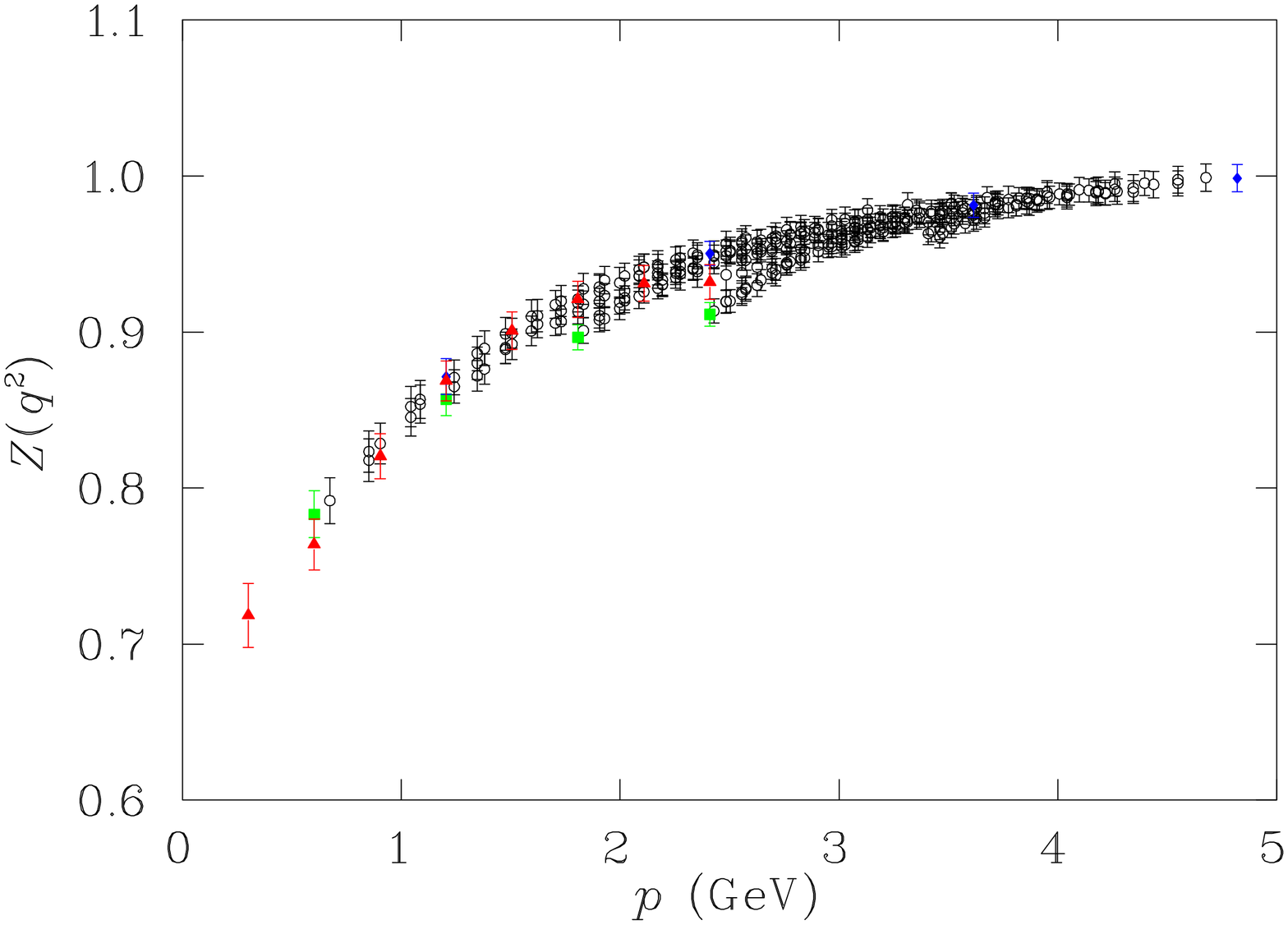,angle=90,width=7cm}
\end{center}
\begin{center}
\epsfig{figure=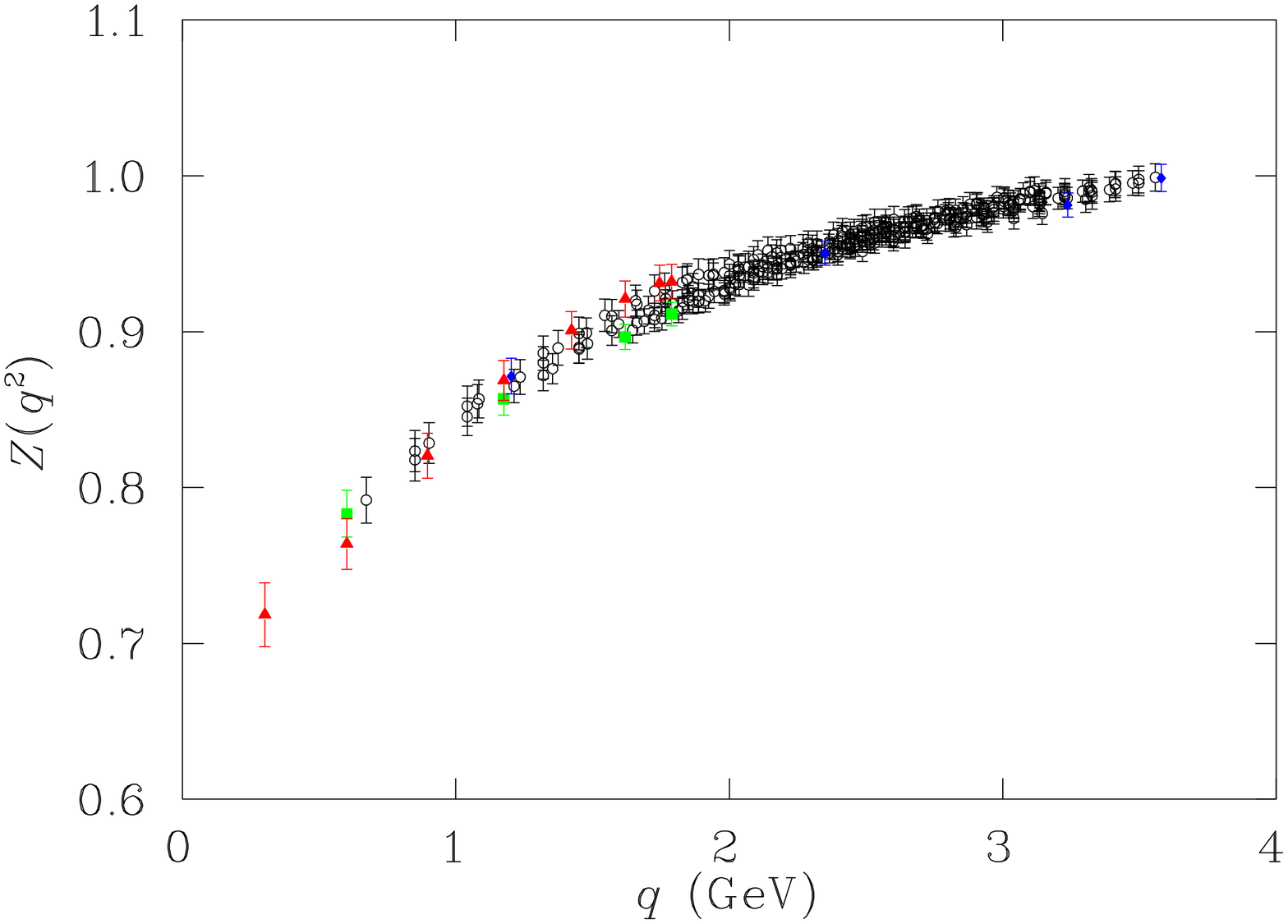,angle=90,width=7cm}
\end{center}
\vspace{-10mm}
\caption{The quark Z function for the Asqtad action ($ma = 0.05$).  When we
use the momentum $q$, defined by the tree-level form of the action, rotational
symmetry breaking is reduced.}
\label{fig:correction}
\vspace{-5mm}
\end{figure}

We investigated the applicaton of tree-level correction 
to the quark propagator by comparing the $Z$ functions using
$p$ and $q$.  One example is shown in Fig.~\ref{fig:correction}.  With the
Asqtad action we see that hypercubic artefacts are small in any case, but
when we use $q$ instead of $p$ they nearly vanish.

It is less clear which momentum variable should be used for the mass function,
because at tree-level it is not multiplied by the momentum, but for 
consistency we use $q$ here as well.  In the case of the mass 
function,  the choice of momentum will actually make little difference to our 
results.

\begin{figure}[tb]
\begin{center}
\epsfig{figure=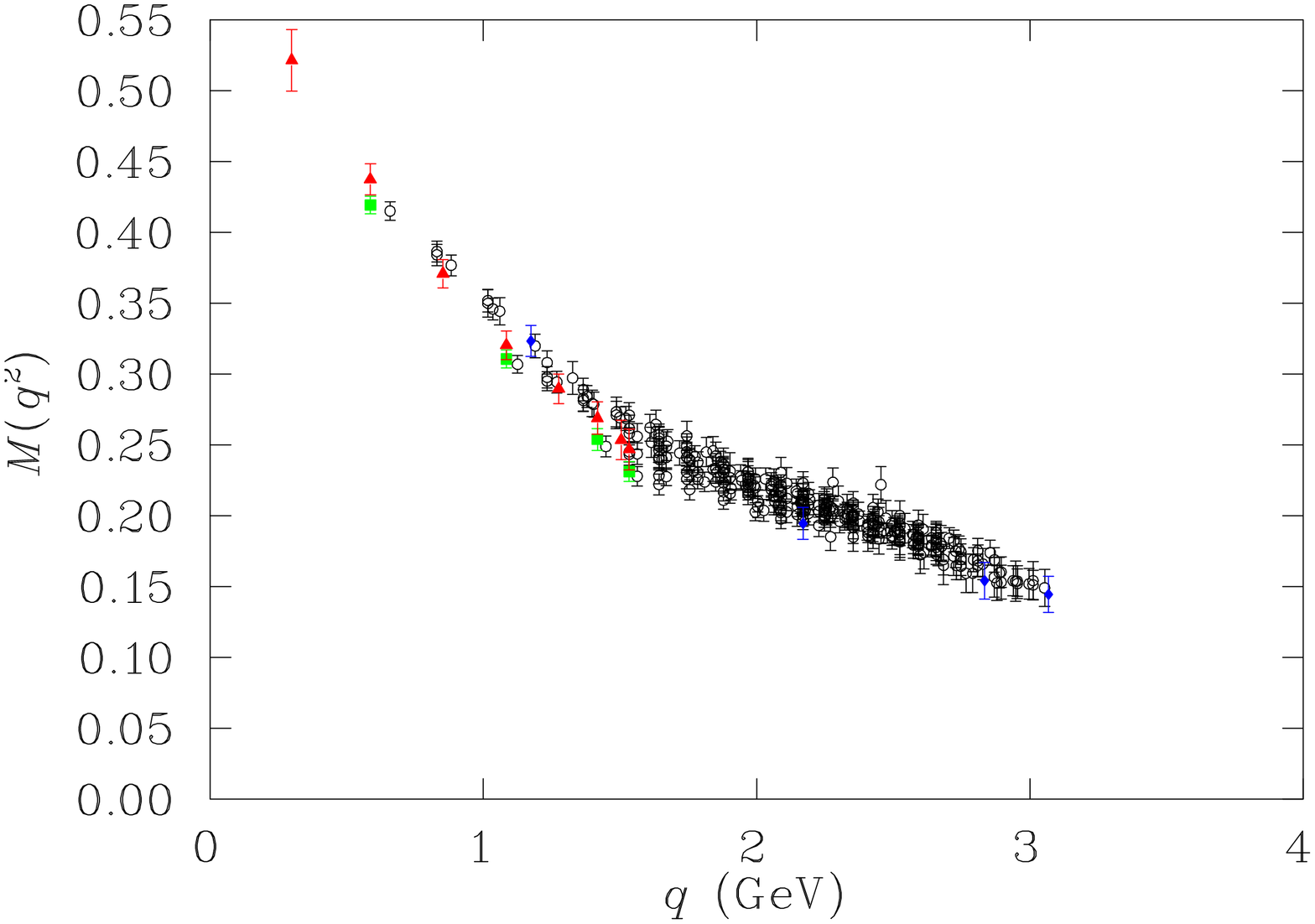,angle=90,width=7cm}
\end{center}
\begin{center}
\epsfig{figure=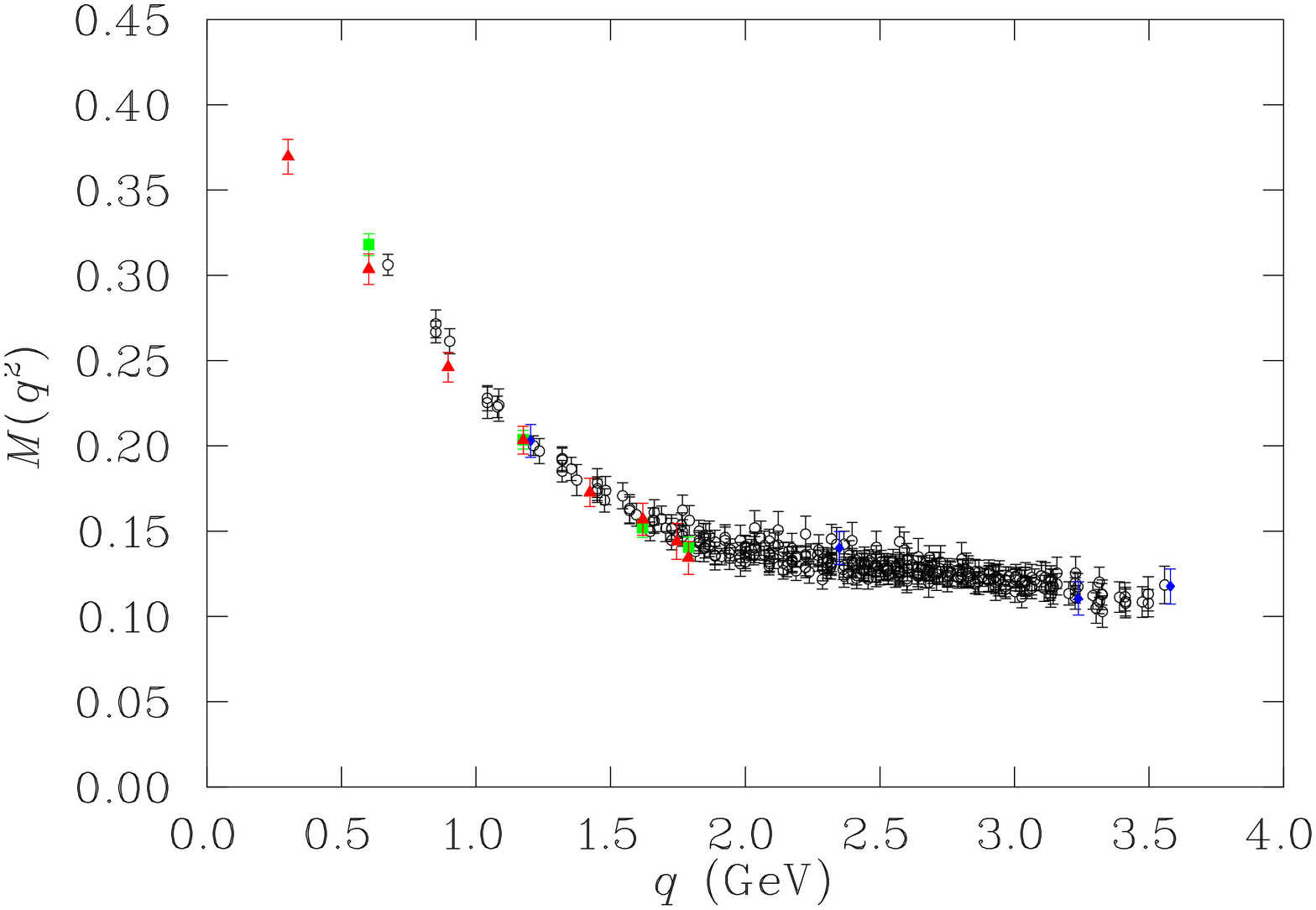,angle=90,width=7cm}
\end{center}
\vspace{-10mm}\caption{Mass function for quark mass $ma = 0.05$ 
($m \simeq 77$ MeV), KS action (top) and Asqtad action (bottom) in Landau 
gauge.}
\label{fig:lan_comp_m05all}
\vspace{-5mm}
\end{figure}

In Fig.~\ref{fig:lan_comp_m05all} the mass function is plotted, in Landau
gauge, for both actions with quark mass $ma = 0.05$.  We see that the KS 
action gives a much larger value for 
M(0) than the Asqtad action and is slower to approach asymptotic behaviour.  
The Asqtad action also shows slightly better rotational symmetry.

The Asqtad action displays clearly
better rotational symmetry in the quark $Z$ function and, curiously, improved
infrared behaviour as well.  The Asqtad action also displays a better approach 
to asymptopia, more smoothly approaching one in the ultraviolet.  Furthermore,
the relative 
improvement increases as the quark mass decreases.  Comparing the mass 
function for the two actions at $ma = 0.0125$, the lowest mass studied here,  
the propagator has much less infrared noise with the 
Asqtad action than with the KS action.

\subsection{Comparison of the gauges}

\begin{figure}[bt]
\begin{center}
\epsfig{figure=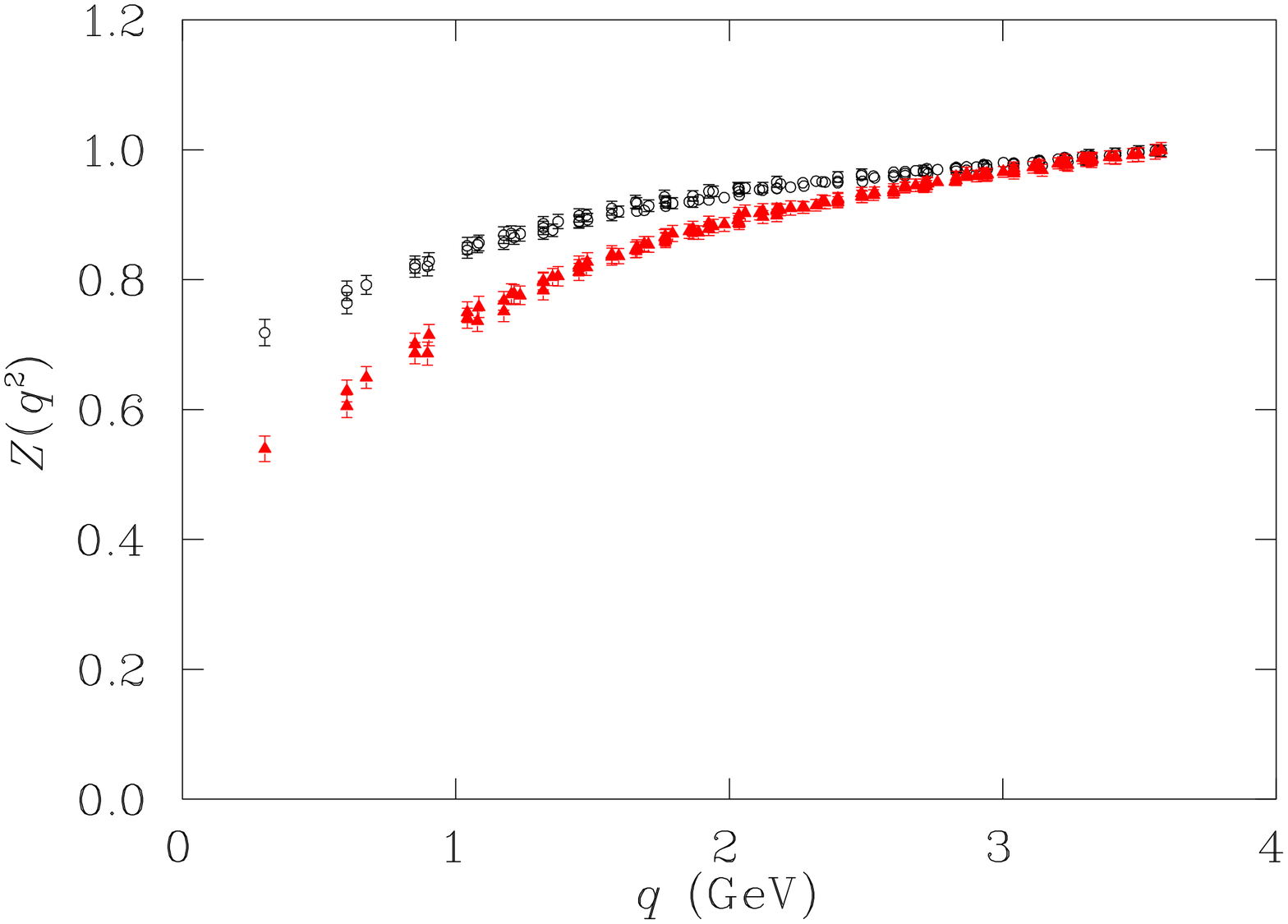,angle=90,width=7cm}
\end{center}
\begin{center}
\epsfig{figure=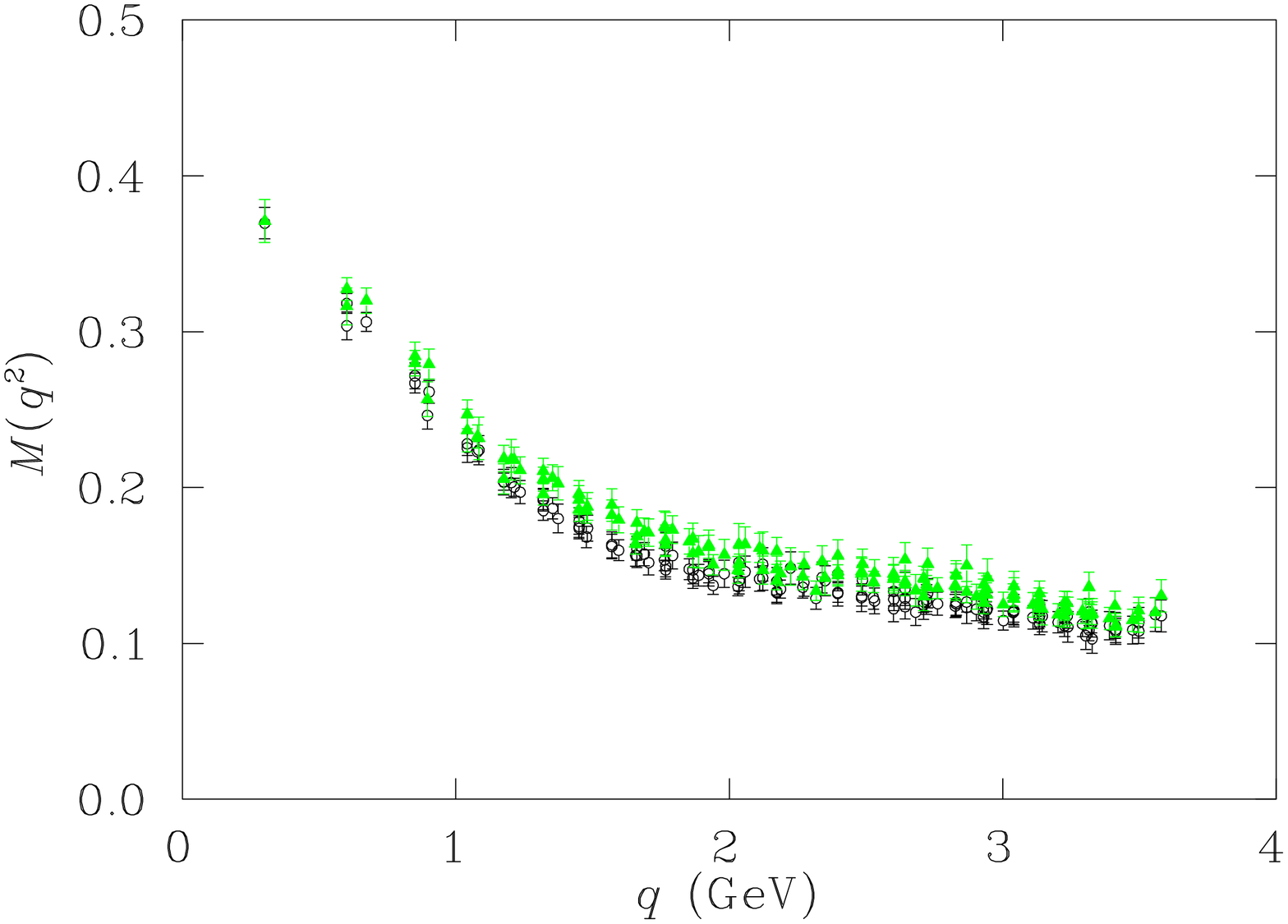,angle=90,width=7cm}
\end{center}
\vspace{-10mm}
\caption{Gauge dependence of the quark Z (top) and mass (bottom) 
functions for the Asqtad action ($ma = 0.05$).  Points marked with open 
circles are in Landau gauge and solid triangles are in \lapI gauge.  
Data has been cylinder cut.}
\label{fig:comp_asq_compm05}
\vspace{-5mm}
\end{figure}

Fig.~\ref{fig:comp_asq_compm05} (top) shows the $Z$ function for the Asqtad
action in Landau and \lapI gauges.  Data has been cylinder cut~\cite{Bon01} 
for easier comparison.  They are in excellent agreement in the
ultraviolet but differ significantly in the infrared.  The dipping of $Z$ in
the infrared is associated with dynamical chiral symmetry breaking.
There appears to be some slight difference in the $Z$ function between \lapI 
and \lapII gauges.

Fig.~\ref{fig:comp_asq_compm05} (bottom) shows the mass function for the 
Asqtad action in Landau and \lapI gauges.   The two mass functions agree in 
the ultraviolet and in the infrared, but the Landau gauge mass function sits 
slightly higher in the intermediate region.  The mass functions are nearly 
identical in \lapI and \lapII gauges. 
We have also found that in Landau gauge the mass function has slightly less 
anisotropy at this lattice spacing.

\begin{figure}[t]
\begin{center}
\epsfig{figure=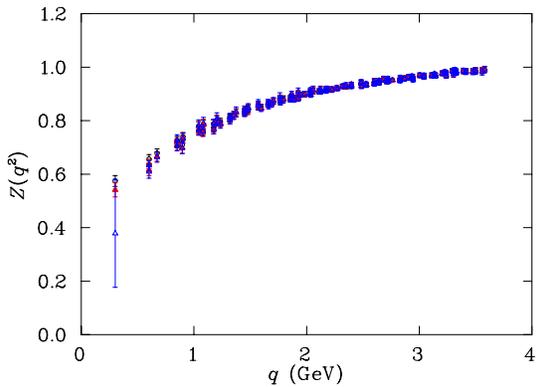,angle=90,width=7cm}
\end{center}
\vspace{-10mm}
\caption{Comparison of the quark Z functions for the three quark masses
$ma = 0.0125$, 0.025 and 0.05,
with the Asqtad action in \lapII gauge.  Data has been cylinder cut.  As in
Landau gauge, they agree to within errors, although there is a systematic
ordering of the infrared points from heaviest quark (top) to lightest 
(bottom).}
\label{fig:lap2_asq_Zcomp}
\vspace{-5mm}
\end{figure}

Landau gauge seems to respond somewhat better than \lapII gauge to vanishing
quark mass.  With the smallest quark mass ($ma = 0.0125$) the lowest momentum
points have large errors in \lapII gauge compared with Landau gauge.  This
can be seen in the $Z$ function in Fig.~\ref{fig:lap2_asq_Zcomp}.

\lapIII performs very poorly.  We found that many of the matrices 
had vanishingly small determinants (compared to numerical precision), which 
destroyed the projection onto $SU(3)$. 
Problems with \lapIII have also been seen in the gluon 
propagator~\cite{Bow01b}.

\subsection{Modelling the quark propagator}
 
\begin{figure}[t]
\begin{center}
\epsfig{figure=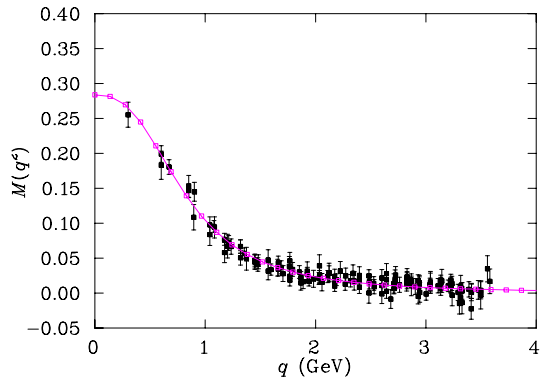,angle=90,width=7cm}
\end{center}
\vspace{-10mm}\caption{Mass function extrapolated to the chiral limit.  
Errors are Jack-knife.  Fit parameters are
c = 0.030(4), $\Lambda$ = 870(60) MeV, $m_0$ = 0.0, $\alpha$ = 1.52(23), 
$\chi^2$ / dof = 0.49.}
\label{fig:lan_asq_chiralfit}
\vspace{-5mm}
\end{figure}

We performed an extrapolation to the chiral limit using a quadratic fit.
The model ansatz 
\begin{equation}
\label{eq:massfit}
M(q) = \frac{c\Lambda^{1+2\alpha}}{q^{2\alpha} + \Lambda^{2\alpha}} + m_0,
\end{equation}
which is a generalisation of the one used in Ref.~\cite{Sku01a}, was then fit
to each mass function.  The resulting fit parameters are shown in 
Table~\ref{tab:lan_fits} and Fig.~\ref{fig:lan_asq_chiralfit}
shows the best fit for the mass function in the chiral limit.  We show here
only results for Landau gauge, as they are compatible with the Laplacian gauge
results, but slightly cleaner.  We find that $\alpha > 1$ is increasingly 
favoured as the quark mass approaches zero.

\begin{table*}[ht]
\caption{Best-fit parameters for the ansatz,~\eref{eq:massfit}, in Landau
gauge, in physical units.  The table is divided into two sections, the first
has $\alpha = 1.0$ fixed and the second leaves $\alpha$ as a free parameter.
For the last fit in each case, the ultraviolet mass, $m_0$, was fixed to zero.
Generally, $\alpha$ is not well determined by the fits, but $\alpha > 1$ seems
to be favoured in the chiral limit.}
\label{tab:lan_fits}
\vspace{2mm}
\begin{tabular}{ccccccc}
\hline
 $m$ (MeV) & c & $\Lambda$ (MeV) & $m_0$ (MeV) & $\alpha$ & M(0) (MeV) & $\chi^2$ /dof \\
\hline
115 & 0.40(2)  & 920(20)  & 144(7)  &    1.0   & 467(9)  & 0.38  \\
 96 & 0.36(5)  & 890(70)  & 118(8)  &    1.0   & 440(20) & 0.42  \\
 77 & 0.41(5)  & 830(70)  &  95(7)  &    1.0   & 430(20) & 0.42  \\
 58 & 0.45(4)  & 770(50)  &  70(7)  &    1.0   & 420(20) & 0.51  \\
 38 & 0.49(6)  & 720(60)  &  44(6)  &    1.0   & 400(30) & 0.56  \\
 19 & 0.54(7)  & 670(60)  &  18(6)  &    1.0   & 380(30) & 0.69  \\
  0 & 0.56(8)  & 660(50)  & -12(6)  &    1.0   & 350(30) & 0.66  \\ 
  0 & 0.77(15) & 530(50)  &  0.0    &    1.0   & 410(40) & 1.3   \\
\hline
115 & 0.28(1)  &1000(30)  & 157(7)  & 1.25(4)  & 433(7)  & 0.38  \\
 96 & 0.28(2)  & 975(40)  & 131(9)  & 1.26(10) & 408(9)  & 0.37  \\
 77 & 0.30(5)  & 940(50)  & 110(10) & 1.29(6)  & 380(10) & 0.36  \\
 58 & 0.30(2)  & 920(40)  &  80(6)  & 1.30(2)  & 360(10) & 0.41  \\
 38 & 0.36(6)  & 830(50)  &  55(7)  & 1.28(7)  & 350(20) & 0.46  \\
 19 & 0.34(4)  & 830(100) &  28(5)  & 1.35(14) & 310(40) & 0.55  \\
  0 & 0.30(4)  & 870(60)  &  0.0    & 1.52(23) & 260(20) & 0.49  \\  
\hline
\end{tabular}
\end{table*}

\section{Conclusions}

We have seen that the lattice quark propagator has better rotational
symmetry and displays more rapid approach to asymptotic behaviour with the 
Asqtad action than with the standard Kogut-Susskind action.
Three implementations of the Laplacian gauge were investigated, and it
was found that \lapI and \lapII gauges gave similar results to Landau gauge.
\lapIII worked very poorly.  The mass function showed very little sensitivity
to the choice of gauge, but some change was seen in the quark Z function.
We attempted to model the mass function and saw that the ansatz provided a 
good fit to the data.

As we have simulated on only one lattice, it remains to do a thorough 
examination of discretisation and finite volume effects.  A natural extension
of the mass function ansatz would be to include the correct asymptotic 
behaviour, and this would require testing on much finer lattices.

\section*{Acknowledgments}

The authors wish to thank Derek Leinweber and Jonivar Skullerud for useful
discussions.  POB and UMH would like to thank the organisers of the Lattice 
Hadron Physics workshop (Cairns, 2001) for a great workshop and POB also 
thanks them for financial support.

%%%%%%%%%%%%%%%%% References %%%%%%%%%%%%%%%

\end{document}